\documentclass[12pt,a4paper]{article}
\usepackage{graphicx}
\usepackage{amsfonts}
\usepackage{amssymb}

\def\hybrid{\topmargin -20pt    \oddsidemargin 0pt
        \headheight 0pt \headsep 0pt
        \textwidth 6.25in       
        \textheight 9.5in       
        \marginparwidth .875in
        \parskip 5pt plus 1pt   \jot = 1.5ex}

\hybrid

\def\baselinestretch{1.2}

\catcode`\@=11

\def\marginnote#1{}
\newcount\hour
\newcount\minute
\newtoks\amorpm
\hour=\time\divide\hour by60
\minute=\time{\multiply\hour by60 \global\advance\minute by-\hour}
\edef\standardtime{{\ifnum\hour<12 \global\amorpm={am}%
        \else\global\amorpm={pm}\advance\hour by-12 \fi
        \ifnum\hour=0 \hour=12 \fi
        \number\hour:\ifnum\minute<10 0\fi\number\minute\the\amorpm}}
\edef\militarytime{\number\hour:\ifnum\minute<10 0\fi\number\minute}

\def\draftlabel#1{{\@bsphack\if@filesw {\let\thepage\relax
   \xdef\@gtempa{\write\@auxout{\string
      \newlabel{#1}{{\@currentlabel}{\thepage}}}}}\@gtempa
   \if@nobreak \ifvmode\nobreak\fi\fi\fi\@esphack}
        \gdef\@eqnlabel{#1}}
\def\@eqnlabel{}
\def\@vacuum{}
\def\draftmarginnote#1{\marginpar{\raggedright\scriptsize\tt#1}}

\def\draft{\oddsidemargin -.5truein
        \def\@oddfoot{\sl preliminary draft \hfil
        \rm\thepage\hfil\sl\today\quad\militarytime}
        \let\@evenfoot\@oddfoot \overfullrule 3pt
        \let\label=\draftlabel
        \let\marginnote=\draftmarginnote
   \def\@eqnnum{(\theequation)\rlap{\kern\marginparsep\tt\@eqnlabel}%
\global\let\@eqnlabel\@vacuum}  }

\def\preprint{\twocolumn\sloppy\flushbottom\parindent 2em
        \leftmargini 2em\leftmarginv .5em\leftmarginvi .5em
        \oddsidemargin -.5in    \evensidemargin -.5in
        \columnsep .4in \footheight 0pt
        \textwidth 10.in        \topmargin  -.4in
        \headheight 12pt \topskip .4in
        \textheight 6.9in \footskip 0pt
        \def\@oddhead{\thepage\hfil\addtocounter{page}{1}\thepage}
        \let\@evenhead\@oddhead \def\@oddfoot{} \def\@evenfoot{} }

\def\numberbysection{\@addtoreset{equation}{section}
        \def\theequation{\thesection.\arabic{equation}}}

\def\underline#1{\relax\ifmmode\@@underline#1\else
        $\@@underline{\hbox{#1}}$\relax\fi}

\def\titlepage{\@restonecolfalse\if@twocolumn\@restonecoltrue\onecolumn
     \else \newpage \fi \thispagestyle{empty}\c@page\z@
        \def\thefootnote{\fnsymbol{footnote}} }

\def\endtitlepage{\if@restonecol\twocolumn \else \newpage \fi
        \def\thefootnote{\arabic{footnote}}
        \setcounter{footnote}{0}}  

\catcode`@=12
\relax

\def\figcap{\section*{Figure Captions\markboth
        {FIGURECAPTIONS}{FIGURECAPTIONS}}\list
        {Figure \arabic{enumi}:\hfill}{\settowidth\labelwidth{Figure
999:}
        \leftmargin\labelwidth
        \advance\leftmargin\labelsep\usecounter{enumi}}}
 \relax
\def\tablecap{\section*{Table Captions\markboth
        {TABLECAPTIONS}{TABLECAPTIONS}}\list
        {Table \arabic{enumi}:\hfill}{\settowidth\labelwidth{Table
999:}
        \leftmargin\labelwidth
        \advance\leftmargin\labelsep\usecounter{enumi}}}
 \relax
\def\reflist{\section*{References\markboth
        {REFLIST}{REFLIST}}\list
        {[\arabic{enumi}]\hfill}{\settowidth\labelwidth{[999]}
        \leftmargin\labelwidth
        \advance\leftmargin\labelsep\usecounter{enumi}}}
 \relax
\makeatletter
\newcounter{pubctr}
\def\publist{\@ifnextchar[{\@publist}{\@@publist}}
\def\@publist[#1]{\list
        {[\arabic{pubctr}]\hfill}{\settowidth\labelwidth{[999]}
        \leftmargin\labelwidth
        \advance\leftmargin\labelsep
        \@nmbrlisttrue\def\@listctr{pubctr}
        \setcounter{pubctr}{#1}\addtocounter{pubctr}{-1}}}
\def\@@publist{\list
        {[\arabic{pubctr}]\hfill}{\settowidth\labelwidth{[999]}
        \leftmargin\labelwidth
        \advance\leftmargin\labelsep
        \@nmbrlisttrue\def\@listctr{pubctr}}}
 \relax
\makeatother
\newskip\humongous \humongous=0pt plus 1000pt minus 1000pt

\newif\ifdtup

\relax

\def\be{\begin{equation}}
\def\ee{\end{equation}}
\def\ba{\begin{eqnarray}}
\def\ea{\end{eqnarray}}

\def\del{\partial}

\def\a{\alpha}

\def\D{\Delta}
\def\e{\epsilon}

\def\th{\theta}

\def\m{\mu}
\def\n{\nu}

\def\l{\lambda}

\def\s{\sigma}

\def\no{\noindent}

\def\qq{\qquad}

\def\IR{\relax{\rm I\kern-.18em R}}

\def \ha {{1\over 2}}

\def \ov {\over}

\def\IR{\relax{\rm I\kern-.18em R}}
\def\inv{^{\raise.15ex\hbox{${\scriptscriptstyle -}$}\kern-.05em 1}}

\begin{document}


\renewcommand{\theequation}{\arabic{equation}}

\newcommand{\beq}{\begin{equation}}
\newcommand{\eeq}[1]{\label{#1}\end{equation}}
\newcommand{\ber}{\begin{eqnarray}}
\newcommand{\eer}[1]{\label{#1}\end{eqnarray}}
\newcommand{\eqn}[1]{(\ref{#1})}
\begin{titlepage}
\begin{center}

\hfill CERN-TH/2000-120\\
\vskip -.1 cm
\hfill NEIP-00-011\\
\vskip -.1 cm
\hfill hep--th/0004148\\

\vskip .5in

{\large \bf An $\mathcal{N} =2$ gauge theory and its supergravity dual}

\vskip 0.4in

{\bf A. Brandhuber}${}^1$\phantom{x}and\phantom{x} {\bf K. Sfetsos}${}^2$ 
\vskip 0.1in
{\em ${}^1\!$Theory Division, CERN\\
     CH-1211 Geneva 23, Switzerland\\
\footnotesize{\tt brandhu@mail.cern.ch}}\\
\vskip .2in
{\em ${}^2\!$Institut de Physique, Universit\'e de Neuch\^atel\\
Breguet 1, CH-2000 Neuch\^atel, Switzerland\\
\footnotesize{\tt sfetsos@mail.cern.ch}}\\

\end{center}

\vskip .3in

\centerline{\bf Abstract}

\noindent
We study flows on the scalar manifold of $\mathcal{N} = 8$ 
gauged supergravity in five dimensions
which are dual to certain mass deformations
of $\mathcal{N} =4$ super Yang--Mills theory. In particular, we consider
a perturbation of the gauge theory by a mass term for the adjoint 
hyper-multiplet, giving rise to an $\mathcal{N} =2$ theory.
The exact solution of the 5-dim gauged supergravity equations of motion 
is found and the metric is uplifted to a ten-dimensional background
of type-IIB supergravity. 
Using these geometric data and the AdS/CFT
correspondence we analyze the spectra of certain operators as well 
as Wilson loops on the dual gauge theory side.
The physical flows are parametrized by a single non-positive constant 
and describe part of the Coulomb branch of the $\mathcal{N} =2$ theory
at strong coupling.
We also propose a general criterion to distinguish between `physical' and
`unphysical' 
curvature singularities. Applying it in many backgrounds 
arising within the AdS/CFT correspondence we find results 
that are in complete agreement with field theory expectations.

\vskip 2 cm
\noindent
CERN-TH/2000-120\\
April 2000\\
\end{titlepage}
\vfill
\eject

\def\baselinestretch{1.2}
\baselineskip 16 pt
\noindent

\section{Introduction}

The AdS/CFT correspondence \cite{malda,gkp,witten} provides a powerful tool
for studying ${\cal N}=4$ supersymmetric Yang--Mills theory in
four dimensions at large $N$ and large 't Hooft coupling.
In particular, there exist precise prescriptions to calculate
correlation functions, spectra of gauge invariant operators,
Wilson loops and $c$-functions in supergravity. These data can be
compared with field theory or provide non-trivial predictions
for strongly coupled field theories. A natural question is
whether this correspondence can be extended to theories with
spontaneously or manifestly broken superconformal symmetry.
Such theories arise either by giving vacuum expectation
values to scalar fields \cite{malda}, \cite{MW}-\cite{KS1} or by
deformations of the conformal
theory with relevant operators \cite{dz1}-\cite{Behrndt}.

These issues can all be treated efficiently 
in the context of five-dimensional
gauged supergravity \cite{PPN,GRW}
and the resulting backgrounds are kink-type solutions
with four-dimensional Poincar\'e invariance which approach AdS
asymptotically.
A related question concerns the uplifting of these solutions
to solutions of type-IIB supergravity/string theory, 
which can be quite involved,
as we will see in this work. It is also of interest for the program
of consistent truncations \cite{TRUN} which, as yet, has not been completed
in the case of ${\cal N}=8$ gauged supergravity in five dimensions.

In this note we will study a supergravity dual
of a particular deformation of the
${\cal N}=4$ SYM theory by a mass term that preserves
${\cal N}=2$ supersymmetry \cite{sw,donagi}.
This is simply ${\cal N} = 2$ supersymmetric gauge theory with gauge group 
$SU(N)$ coupled to a massive
hypermultiplet in the adjoint representation of the gauge group. 
One can think of this model also as 
${\cal N} = 1$ supersymmetric QCD with three chiral multiplets 
$\Phi_{i=1,2,3}$ in the adjoint, 
out of which, one is massless and the other two have equal masses.
Other choices of the massterms are of course possible and lead to
models with ${\cal N} = 1$ supersymmetry. Such models have 
been studied previously in the context of the AdS/CFT correspondence 
\cite{gppz2,Dorey,polstr}.

The outline of this paper is as follows: in section 2 we present
the background of gauged supergravity that is dual to 
${\cal N} = 4$ SYM deformed by a mass term for one hypermultiplet.
We find a family of solutions that is parametrized by one real 
constant $c$, whose value determines the physics and 
study fluctuation in this background.
For $c\leq 0$ it turns out that we describe 
the strong coupling regime of part of 
the Coulomb branch of an ${\cal N}=2$ theory, whereas flows with $c>0$
are unphysical.
In section 3, we compute the uplifted metric in ten dimensions. With the
help of this type-IIB background we calculate expectation values
of Wilson loops that correspond to the potential of an external heavy
quark-antiquark pair.
Finally, in section 4 we present some concluding remarks and comment on 
the singularities that are typical for
backgrounds of non-conformal theories.

\no
{\bf Note added}

In the final stages of our work, the paper \cite{PW2} appeared
that has considerable overlap with ours. 
In addition to the uplifted ten-dimensional  
metric, that we also computed independently, these authors presented
the axion/dilaton and the complex two-form in ten dimensions.
A numerical investigation has also appeared before in \cite{gubser}.

\section{A dual of ${\cal N} = 2$ supersymmetric gauge theories}

Our starting point is the action of five-dimensional gauged supergravity
\be\label{action}
S = \int d^5x \left\{ \frac{1}{4} \mathcal{R} - \frac{1}{2} \sum_{I} 
\partial_{\mu} \a_I \partial^{\mu} \a_I - P(\a) \right\}\ ,
\ee
where we have chosen canonical kinetic terms
for the scalars. This is possible only in certain sub-sectors of the full 
$E_{6(6)}/USp(8)$ coset space sigma-model
parameterizing the 42 scalars of the theory. We have also set to zero all 
other tensor fields.

For the applications we have in mind we need the scalars that correspond
to dimension 2 and 3 operators, which are in the ${\bf 20}$ and
${\bf 10}$ representation of $SO(6) \sim SU(4)$, respectively.
The massless 5-dim dilaton and axion will not play a r\^ole and are constant.
They correspond to exactly marginal deformations in the gauge theory.
Furthermore, we are interested in solutions of (\ref{action}) that preserve
part of the supersymmetries. For such supersymmetric flows it is known
that the scalar potential $P$  can be written in terms of a 
{\it superpotential}, which we will denote as $W(\a)$:
\be\label{superpot}
P(\a) = \frac{1}{8} \sum_I \left( \frac{\partial W}{\partial \a_I} \right)^2
- \frac{1}{3} W^2\ .
\ee
Furthermore, we demand that the solutions of (\ref{action}) preserve
four-dimensional Poincar\'e invariance along the brane directions,
and that the metric asymptotes the $AdS_5$ space-time near the boundary
which we take to be at $r \to \infty$. For the metric we make the
ansatz
\be\label{metric}
ds_5^2 = e^{2 A(r)} dx_{||}^2 + dr^2\ ,
\ee
or, if we need the metric in its conformally flat version then
\be\label{metricv2}
ds_5^2 = e^{2 A(z)} \left( dx_{||}^2 + dz^2 \right) \ ,
\ee
where the relation between the different coordinate choices is
$dr = - e^A dz$. The boundary at $r=\infty$ corresponds to $z=0$.

On the supergravity side two scalars, denoted by $\a_2$ and $\a_3$, 
are involved and belong to the ${\bf 20}$ and to the ${\bf
10 + \overline{10}}$ representation of $SO(6)$, respectively. 
They are dual to
operator-bilinears in scalars and fermions as:
\ba\label{operators}
&& \a_2 : \qq \mathcal{O}_2  =  \mathrm{Tr}
\left( \bar{Z}_1 Z_1 + \bar{Z}_2 Z_2 -2  \bar{Z}_3 Z_3 \right)\ ,
 \nonumber \\
&& \a_3:\qq  \mathcal{O}_3  =  \mathrm{Tr}
\left( \l_1 \l_1 + \l_2 \l_2 + \ldots + h.c. \right) \ ,
\ea
where the $Z_{i=1,2,3}$ denote the complex scalar components of the
chiral superfields $\Phi_i$, the $\l_{i=1,2,3}$ are the corresponding 
fermionic components, 
and the $\ldots$ denote scalar trilinear terms.
The non-vanishing scalar $\a_2$ reduces the gauge group 
to $SU(2) \times SU(2) \times U(1) \subset SU(4)$ under which
the scalars in the ${\bf 6}$ and the fermions in the ${\bf 4}$ decompose as:
\ba
{\bf 6} & \to & (1,1)_2 + (1,1)_{-2} + (2,2)_0 \ ,
\nonumber \\
{\bf 4} & \to & (2,1)_{1} + (1,2)_{-1} \ .
\ea
We see that the $U(1)$ is to be identified with
the $U(1)_R$ symmetry of the field theory, since the two
scalars in the vector multiplet correspond to the two
$SU(2)$ singlets in the ${\bf 6}$
which have charge $\pm2$ under the $U(1)$ as in field theory, 
whereas the scalars of the hypermultiplet are singlets. 
The scalar $\a_3$ breaks one of the $SU(2)$'s to $U(1)$, and the
unbroken one can be identified with the $SU(2)_R$ symmetry of
the gauge theory under which the scalars in the vector multiplet
are singlets and the scalars in the hypermultiplet are doublets.
Furthermore, the two fermions in the hypermultiplet are singlets
under $SU(2)_R$ and have $U(1)_R$ charge $-1$, the two fermions
in the vector multiplet are $SU(2)_R$ doublets and their $U(1)_R$
charge is $+1$. So in our decomposition the second $SU(2)$ is
broken to $U(1)$.

The potential in the scalar field space is obtained by a truncation of the 
four scalar potential that has been computed in \cite{khavaev} and reads
\ba\label{potentials}
P  & =& \frac{1}{16} e^{-\frac{4}{\sqrt{6}} \a_2}\left( 
-4-8 e^{\sqrt{6} \a_2} \cosh \sqrt{2} \a_3 + e^{2 \sqrt{6} \a_2} \sinh^2
\sqrt{2} \a_3 \right) \ ,
\nonumber \\
W & = & -e^{-2 \a_2 / \sqrt{6} } - 
\frac{1}{2} e^{4 \a_2 / \sqrt{6} } \cosh \sqrt{2} \a_3\ ,
\ea
where $W$ is the corresponding superpotential (cf. \eqn{superpot}).
For a supersymmetric flow the equations of motion of (\ref{action}) reduce
to a set of first order equations:
\ba\label{eoms}
\dot{\a}_2 & = & \frac{1}{2} \partial_{\a_2} W = \frac{1}{\sqrt{6}}
e^{- 2 \a_2/ \sqrt{6} } - 
\frac{1}{\sqrt{6}} e^{4 a_2 /\sqrt{6} } \cosh \sqrt{2} \a_3\ ,
\nonumber \\
\dot{\a}_3 & = & \frac{1}{2} \partial_{\a_3} W = 
- \frac{1}{2 \sqrt{2}} e^{4 \a_2 / \sqrt{6}} \sinh \sqrt{2} \a_3\ , \\
\dot{A} & = & -\frac{1}{3} W\ ,
 \nonumber
\ea
where the dot denotes the derivative with respect to the variable $r$ 
introduced in (\ref{metric}). 
As a consistency check, we note that
we may set the scalar $\a_3$ to zero, 
as it is obvious from \eqn{eoms}. Then, the 
remaining scalar describes the part of the Coulomb branch of the 
${\cal N}=4$ SYM theory corresponding to the background of D3-branes
distributed on a disc and $SO(4) \times U(1)$ symmetry.\footnote{There 
should be a generalization of our discussion so far that includes 
a third scalar that further breaks $SU(2)\times U(1)\times U(1)$ to
$SU(2)\times U(1)$. When the scalar in the $\bf 10$ is turned off 
this should be the part of the Coulomb branch of the 
${\cal N}=4$ SYM theory corresponding to the background of D3-branes
distributed on a ellipsoid, which has an $SO(4) $ symmetry. This solution was 
constructed in \cite{bakas1}.}

The systems of equations in (\ref{eoms}) form a rather complicated set 
of coupled
first order differential equations, which, at first sight, do not
have a solution that can be written down in a closed form. This is true 
for the coordinate $r$, but if we take the scalar $\a_3$ as a new radial 
coordinate\footnote{From 
(\ref{chaa}) and the remark after \eqn{metricv2} it is clear 
that $\a_3$ is a monotonously decreasing function of $r$.} 
a solution can be given in terms of elementary functions. Note that
the equations in \eqn{eoms} have a ${\bf Z}_2$ symmetry which transforms 
$\a_3 \to - \a_3$, so that we may restrict our analysis to $\a_3 \geq 0$ 
without loss of generality. For the scalar field $\a_3$ we find 
\be
e^{\sqrt{6} \a_2}  =  \sinh^2 \sqrt{2} \a_3 \left[ c +
\ln \tanh \left(\frac{\a_3}{\sqrt{2}} \right) \right] +
\cosh \sqrt{2} \a_3 \ ,
\label{scasol}
\ee
where $c$ is an integration constant. As we will see the physical picture
depends crucially on whether $c$ is positive, negative or zero.

\begin{figure}[h]
\begin{center}
\includegraphics[width=0.5\textwidth]{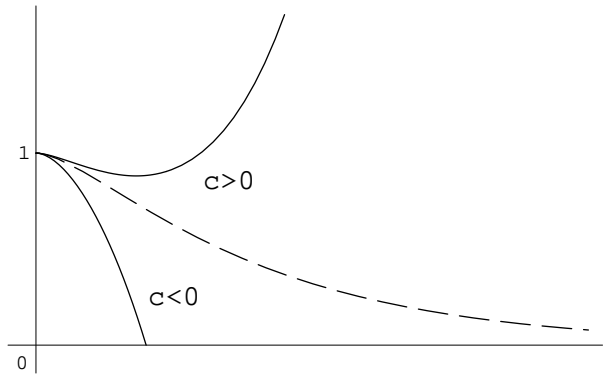}
\caption{This figure shows the different behaviours of the 
function $e^{\sqrt{6} \a_2}$ as a function of $\a_3$ using \eqn{scasol}. 
For $c<0$, it decreases monotonously to zero at a finite 
value $\a_3^{\rm max}$. For $c=0$ (the dashed curve)
the same minimum at zero is attained, 
but for $\a_3\to \infty$. In contrast, for $c>0$ the function first 
reaches a minimum, at a finite value $\a_3^{\rm min}$, before it 
runs off to infinity at $\a_3\to \infty$. }
\end{center}
\end{figure}
For the 5-dim metric we find
\be\label{solution}
ds_5^2  =  \frac{1}{\sinh^2 \sqrt{2} \a_3} \left( 2^{2/3} e^{4\a_2/\sqrt{6}}
dx_{||}^2 + 8 e^{-8\a_2/\sqrt{6}} d\a_3^2 \right)\ .
\ee
The transition to the conformally flat metric (\ref{metricv2}) 
with conformal factor
\be
e^{2 A(z)} = {2^{2/3} e^{4\a_2/\sqrt{6}}\ov \sinh^2 \sqrt{2} \a_3}\ ,
\label{dj2}
\ee
is given by the relation of the differentials 
\be
dz = 2^{7/6} e^{-\sqrt{6} \a_2} d\a_3\ .
\label{chaa}
\ee
We note that, for generic values of the constant $c$,
the explicit dependence of $\a_3$ on $z$ cannot be obtained 
from \eqn{chaa}, since the corresponding integral cannot be explicitly
evaluated.

Let us consider the behaviour near the boundary where
the background becomes $AdS_5$.
The boundary is approached as $\a_3 \to 0$ which in turn, using \eqn{scasol},
forces $\a_2 \to 0$ as well. In this limit we find $z \simeq 2^{7/6} \a_3$.
Furthermore, we want to check that the scalars behave as expected from
the AdS/CFT correspondence. The scalar $\a_3$ is dual to the fermion
bilinear operator $\mathcal{O}_3$. Our field theory is
obtained as a mass deformation of the $\mathcal{N}=4$ theory and therefore 
the solution
near the horizon should behave as a pure mass term, i.e. $\a_3 \sim z^1$
and there should be no $z^3$ term corresponding to a gluino condensate.
On the other hand, the behaviour of the scalar $\a_2$ is a beautiful example 
of the double
r\^ole that the scalars play in the AdS/CFT correspondence --- they can 
appear as deformations
of the conformal theory, or they parameterize states in the theory.
Here we expect both since we have to give mass to the scalar components
of the chiral multiplets $\Phi_{1,2}$ as dictated by supersymmetry,
and the gauge theory has a Coulomb branch parametrized by the vev of
$Z_3$, i.e. $\a_2$ should have contributions of the type 
$z^2$ and $z^2 \log z$.
From our explicit solution we immediately find that
\ba\label{asympt}
\a_3 & \sim &  z \nonumber \\
\a_2 & \sim &  \frac{1- \ln 2 + 2 c}{\sqrt{6}} z^2  +
\frac{2}{\sqrt{6}} z^2 \ln z\ .
\label{gjk3}
\ea
From these expressions we see that the correct terms appear, as expected,
and that there is a one parameter family of solutions labelled 
by $c$.\footnote{As was noted in \cite{gubser}, there is a critical line 
in the space of solutions corresponding to $c=0$, in our notation.
In the notation of \cite{gubser}, this gives $b_{\rm cr}=
(1 - \ln 2)/\sqrt{6}\simeq 0.125$, thus confirming the critical value
found numerically in \cite{gubser}.}

In this paper we will also be interested in the spectrum corresponding 
to the massless scalar equation in a background metric of the type 
\eqn{metricv2} and with a plane wave ansatz along the flat coordinates 
denoted by $x_{||}$. Within the AdS/CFT correspondence 
\cite{malda,gkp,witten} the solutions and eigenvalues 
of the massless scalar equation have been associated, 
on the gauge theory side, with the spectrum of the operator
${\rm Tr} F^2$, whereas those of the graviton fluctuations 
polarized in the directions parallel to the brane,
with the energy momentum tensor $T_{\m\n}$ 
\cite{gkt,gkp,witten}.
Though a priori not to be expected, these two spectra and the 
corresponding eigenfunctions coincide \cite{brand2} (for a more recent 
discussion see \cite{DeWolfe}).
It is well known that the entire analysis can be cast into an 
equivalent Schr\"odinger problem with potential 
\be
V = \frac{9}{4} {A^\prime}^2 + \frac{3}{2} A^{\prime\prime}\ 
\label{schro}
\ee
and eigenvalue equal to the mass squared ($M^2$).
Since, as already noted, for generic values of the constant $c$
we cannot find the explicit dependence of $\a_3$ on $z$, it is not possible
to explicitly evaluate the potential using \eqn{dj2}. 
However, one easily can show that it
is a monotonously decreasing function of $z$.
As in \cite{fgpw2,brand2,bakas1} the potential \eqn{schro} 
has the same form as the potentials
appearing in supersymmetric quantum mechanics and, therefore, the spectrum is 
bounded from below. Its supersymmetric partner is given by an equation similar
to \eqn{schro}, but with a relative minus sign between the two terms. 

\subsection{$c < 0$}

In the following we will study the case $c < 0$ in more detail. In
particular, we will determine the spectrum of fluctuations of a minimal
scalar in this background. 

First, note that $e^{\sqrt{6} \a_2}$ is a monotonically decreasing
function of $\a_3$ and, as it turns out, it has only a finite
range (see fig. 1)
\be
0 \leq \a_3 \leq \a_3^{\mathrm{max}} ~~.
\label{jash1}
\ee
In certain limits $\a_3^{\mathrm{max}}$ can be found analytically:
\be
\a_3^{\mathrm{max}} = \left\{
\begin{array}{ll}
-\frac{1}{3 \sqrt{2}} \ln \left(-\frac{3 c}{16}\right) \ , & c \to 0^-\ ,
 \\
\frac{1}{\sqrt{-2 c}} \ , & c \to - \infty  \ .
\end{array} 
\right.
\label{limi16}
\ee
Using \eqn{scasol} and \eqn{chaa} we find that
\be
e^{\sqrt{6} \a_2} \simeq \frac{2 \sqrt{2}}{\sinh \sqrt{2} \a_3^{\mathrm{max}}}
\left(\a_3^{\mathrm{max}} - \a_3 \right) \ , \qq {\rm as}\quad
\a_3 \to \a_3^{\mathrm{max}} 
\label{gh7}
\ee
and that 
\be z\simeq -  2^{-1/3} \sinh\sqrt{2} \a_3^{\rm max}
 \ln \left( \a_3^{\mathrm{max}} - \a_3 \right)\ ,\qq {\rm as}\quad
\a_3 \to \a_3^{\mathrm{max}} \ .
\label{djh2}
\ee
Therefore $z$ takes values in the whole semi-infinite real line, i.e.
$0\leq z < \infty$. 
From \eqn{dj2}, \eqn{gh7} and \eqn{djh2} we find that 
\be
A \simeq - { 2^{1/3} z\ov 3 \sinh \sqrt{2} \a_3^{\rm max}}\ ,
\qq {\rm as}\quad z\to \infty \ .
\label{gg2}
\ee
From \eqn{schro} one concludes that the 
mass spectrum is continuous with a mass gap. The value of the latter is
found by evaluating the potential for $z \to \infty$. The result is
\be
M^2_\mathrm{gap} = \frac{1}{2^{4/3} \; \sinh^2 \sqrt{2}\a_3^{\mathrm{max}} }
\label{mmggp}\ ,
\ee
and in the two limits corresponding to \eqn{limi16}
\be
M^2_\mathrm{gap} = \left\{
\begin{array}{ll}
\frac{1}{4} \left(-3 c \right)^{2/3} \; , & c \to 0^- \ ,\\
2^{-4/3} (- c) \; , & c \to - \infty\ .
\end{array} 
\right. 
\label{maass1}
\ee
Hence, the mass gap vanishes as $c$ tends to zero and grows large as $c$
approaches large negative values.

It is also worth noting that certain solutions corresponding to points
on the Coulomb branch of $\mathcal{N}=4$ SYM, which were obtained
previously in the literature, can be obtained in special limits of
our solution. The limit $c \to -\infty$ corresponds to a distribution
of D3-branes on a disc of radius 2, in our units. In order to see that
let us change variables as 
\be
\a_3 = \frac{1}{\sqrt{-2 c}} \tanh \left( z/2 \right)\ ,
\ee
and then send $c \to -\infty$. As we can see from \eqn{limi16}, this 
effectively shrinks the range of $\a_3$
to zero. After some elementary algebra we find
\ba
ds_5^2 &=& \frac{\cosh^{2/3}(z/2)}{\sinh^2(z/2)}\left( dz^2 + 2^{2/3} (-c)
dx_{||}^2 \right) ~,
\nonumber \\
e^{\sqrt{6} \a_2} &=& \frac{1}{\cosh^2(z/2)} ~.
\label{metdisc}
\ea
This model has been studied on its own and also exhibits a mass gap.
In fact in order to compute it using our formulae we first re-scale 
the energies by the factor $2^{2/3} (-c)$ as it is evident from the 
expression for the metric in \eqn{metdisc}. Then, the second line in 
\eqn{maass1} gives $M^2_{\rm gap}= 1/4$ which is in agreement with the 
result for the mass gap found in \cite{fgpw2,brand1}, when it is 
expressed in our units.
We finally note that, from a physical point of view a continuous spectrum 
with a mass gap should be  
associated with a complete screening of charges in a quark-antiquark pair
at a finite separation distance (inversely proportional to $M_{\rm gap}$).
This will be further discussed in subsection 3.1.

\subsection{$c>0$}

Contrary to the previous case, $e^{\sqrt{6} \a_2}$ is now not a monotonic
function (see fig. 1). 
For increasing $\a_3$ it first decreases, takes a minimum
at $\a_3 = \a_3^\mathrm{min}$ with
\be
e^{\sqrt{6} \a_2}\Big|_{\rm min} = {1\ov \cosh \a_3^{\rm min}}\ ,
\ee
and then increases monotonically.
The coordinate $z$ on the other hand can easily be seen to
have finite range:
\be
0 \leq z \leq z_0 = 2^{7/6} \int_0^\infty e^{-\sqrt{6} \a_2}  d\a_3\ .
\label{ghe1}
\ee
In certain limits one can work out the value for $\a_3^\mathrm{min}$:
\be
\a_3^\mathrm{min} \simeq \left\{
\begin{array}{ll}
\frac{1}{6 \sqrt{2}} \ln \frac{8}{3 c} \ , & c \to 0^+\ , \\
\sqrt{2} e^{-c -1} \; , & c \to + \infty\ .
\end{array} 
\right. 
\ee
We also obtain from \eqn{scasol} and \eqn{chaa} that 
\be
e^{\sqrt{6} \a_2} \simeq {c\ov 4} e^{2 \sqrt{2}\a_3} \ ,\qq {\rm as}\quad 
\a_3\to \infty\ 
\label{gh3}
\ee
and that
\be e^{-2 \sqrt{2} \a_3} \simeq 2^{-5/3} c (z_0-z)\ ,\qq {\rm as} \quad
z\to z_0^-\ .
\label{gh4}
\ee
Then from \eqn{dj2}, \eqn{gh3} and \eqn{gh4} the
warp factor of the metric becomes
\be
e^A \simeq 2^{7/18} c^{1/6} (z_0 - z)^{1/6}\ , \qq {\rm as} \quad
z\to z_0^-\ ,
\ee
which corresponds to a naked singularity. It turns out to be
a {\it bad} singularity, but we will return to this point later
and discuss it in more detail. The spectrum of scalar fluctuations 
is now discrete since the range of $z$ is finite.
Knowing the warp factor we can easily determine the Schr\"odinger
potential in the limit $z \to z_0^-$. The result is:
\be
V \simeq -\frac{3}{16} \frac{1}{(z-z_0)^2} ~, \qq {\rm as} \quad
z\to z_0^-\ . 
\ee
A WKB calculation yields for the spectrum
\be
M_n^2 =\frac{\pi^2}{z_0^2} m (m+1) + {\cal {O}}(m^0)\ ,\qq n=1,2,\ldots
\label{fdjh2}
\ee
For $c \to \infty$ the geometry reduces to that of a 
distribution of D3-branes smeared on a three-sphere of radius 2 in our units.
In order to illustrate that, similarly to before, 
we first change variables as 
\be
\a_3 = \frac{1}{\sqrt{2 c}} \tan \left( z/2 \right)\ ,
\ee
and then let $c \to \infty$ with the result
\ba
ds_5^2 & = & \frac{\cos^{2/3}(z/2)}{\sin^2(z/2)}\left( dz^2 + 2^{2/3} c
dx_{||}^2 \right) \ ,
\nonumber\\
e^{\sqrt{6} \a_2} & =& \frac{1}{\cos^2(z/2)} ~.
\label{metsph}
\ea
In the limit $c\to \infty$ we find using \eqn{ghe1} that the maximum 
value of $z$ behaves as $z_0\simeq \pi/(2^{1/3} c^{1/2})$. 
Then, after rescaling of the energies by the factor $2^{2/3} c$, the WKB mass
spectrum is just $M^2_{\rm gap}= m(m+1)$ and is in fact the exact result (as 
noted in \cite{fgpw2,brand1}), in the units that we are using.

\subsection{$c=0$}

In this case the ranges of $z$ and $\a_3$ are infinite and
$e^{\sqrt{6} \a_2}$ is a monotonously decreasing function (see fig. 1).
In particular, from \eqn{scasol} and \eqn{chaa} we find that 
\be
e^{\sqrt{6} \a_2} \simeq {4\ov 3} e^{-\sqrt{2} \a_3}\ , \qq {\rm as} \quad
\a_3\to \infty\ 
\label{dj22}
\ee
and that
\be
e^{\sqrt{2} \a_3} \simeq 
{2^{4/3}\ov 3} z \ ,\qq {\rm as} \quad z\to \infty\ .
\label{dj24}
\ee
Hence, 
\be
e^A \sim z^{-4/3}\ ,\qq {\rm as} \quad z\to \infty\ 
\ee
and the Schr\"odinger potential becomes $V \simeq 6/z^2$ for $z\to \infty$.
Therefore the spectrum is continuous and has no gap.
Accordingly, we expect that the screening of charges in a quark-antiquark pair
will be perfect only at an infinite separation. This will be confirmed in 
subsection 3.1.

\section{The lift to ten dimensions}

Recently, it has been shown how general solutions of 5-dim gauged
supergravity can be uplifted to type-IIB supergravity. This is part
of the program to prove that a consistent truncation exists.
This has not been shown for all fields, but for the metric
the full non-linear KK ansatz has been conjectured \cite{khavaev} and it has 
passed several non-trivial tests.

The inverse of the deformed metric on $S^5$ is given in term of the
$27$-bein $\mathcal{V}$ which has global $E_{6(6)}$ and local
$USp(8)$ indices. In the $SL(6,{\bf R}) \times SL(2,{\bf R})$ basis
the vielbein is decomposed as $\mathcal{V} \to 
\left( \mathcal{V}^{IJab}, {\mathcal{V}_{I\a}}^{ab} \right)$
in terms of which the inverse metric becomes 
\be
\hat{g}^{mn} = \Delta^{-2/3} g^{mn} = 2 K^m_{IJ} K^n_{KL}
\tilde{\mathcal{V}}_{IJab} \tilde{\mathcal{V}}_{KLcd}
\Omega^{ac} \Omega^{bd} ~,
\ee
with $\tilde{\mathcal{V}}$ being 
the inverse vielbein and 
$\Delta^2 = \mathrm{det}(g_{mn})/\mathrm{det}(g^{(0)}_{mn})$, where
$(g^{(0)}_{mn})$ is the undeformed five-sphere metric.
Consequently, the ten-dimensional metric takes
the form of a warped product space
\be\label{metricansatz}
ds_{10}^2 = \Delta^{-2/3} ds_5^2 + g_{mn} dy^m dy^n =
\Delta^{-2/3} \left( ds_5^2 + \hat{g}_{mn} dy^m dy^n \right) ~.
\ee
Then using the parameterization presented in \cite{cvetic}
we compute the 5-dim metric to be
\ba
d\hat s^2 & = & {e^{-{2\ov \sqrt{6}}\a_2} \ov \cosh \sqrt{2} \a_3} d\th^2 
+ {e^{-{8\ov \sqrt{6}}\a_2} \sin^2\th \ov \D_1}d\phi_1^2 
\nonumber\\
&& + e^{-{2\ov \sqrt{6}}\a_2} \cos^2\th
\left( {1 \ov \D_1 \cosh \sqrt{2}\a_3 }\s_3^2  + {1\ov \D_2} 
(\s_1^2 + \s_2^2) \right)\ ,
\label{sjd1}
\ea
where
\ba
\D_1 & =& e^{-\sqrt{6} \a_2} \cos^2 \th \cosh \sqrt{2} \a_3 
+ \sin^2\th\ , \nonumber\\
\D_2 &= & e^{-\sqrt{6} \a_2} \cos^2 \th + \sin^2 \th \cosh \sqrt{2} \a_3\ . 
\ea
The Maurer--Cartan forms $\s_i$, $i=1,2,3$ for $SU(2)$ are defined to obey 
$d\s_i = -\e_{ijk} \s_j \wedge \s_k$. A convenient parameterization in 
terms of three Euler angles $\phi_2$, $\phi_3$ and $\psi$ is given by
\ba
\s_1 & = &\ha
 \left(\cos \phi_2 d\psi + \sin \phi_2 \sin \psi d\phi_3 \right)\  ,
\nonumber\\
\s_2 & = & \ha \left(
 - \sin \phi_2 d\psi + \cos \phi_2 \sin \psi d\phi_3 \right)\ ,
\label{maucar}\\
\s_3 & = & \ha \left( d\phi_2  + \cos \psi d\phi_3 \right) \ .
\nonumber 
\ea
Finally, the warp factor in \eqn{metricansatz} is easily computed to be
\be
\Delta^{-2/3} = e^{{2\ov \sqrt{6}} \a_2}
\left(\D_1 \D_2  \cosh \sqrt{2} \a_3 \right)^{1/4}\ .
\ee
The metric \eqn{sjd1} is manifestly $SU(2)$ invariant being written in terms
of the corresponding Maurer--Cartan forms. In addition, there is an
$U(1) \times U(1)$ invariance corresponding to the isometry with respect to the
commuting Killing vectors $\del/{\del\phi_1}$ and $\del/{\del\phi_2}$.

\subsection{Wilson loops and screening}

We now calculate string probes in the ten-dimensional background
\eqn{sjd1} that
correspond to Wilson loops in field theory associated with
the potential between an external quark-antiquark pair \cite{wilson1,wilson2}.
The relevant probe action is that of a fundamental string
which involves only the string frame metric and the NS $B$-field.
As in a similar computation in \cite{brand1}, involving a rotating D3-brane,
we consider trajectories with constant values for the
angles $\th$ and $\phi_1$, as well as for the Euler angles entering into the 
definition of the Maurer--Cartan forms. For such constant values
the pull-back of the $B$-field, which was recently calculated
\cite{PW2}, vanishes, so that the action consists only of the Nambu--Goto
term. 
Since there is an explicit dependence of the metric on $\th$,
consistency requires that the variation of this action with respect to $\th$ is
zero, in order for the equations of motion to be obeyed. It is easy to see that
this procedure allows $\th=0$ or $\th=\pi/2$. 
%

The action can be written as
\be
S_{NG} = \frac{T}{2 \pi \alpha^\prime}\int dx \sqrt{g(\a_3) \left( 
\frac{d\a_3}{dx} \right)^2 + f(\a_3) }\ ,
\ee
with $T$ coming from the trivial integration in the Euclidean-time direction
and $x$ is one of the spatial directions along the brane.
The functions $g(\a_3,\theta) = g_{\tau\tau} g_{\a_3 \a_3}$ , 
$f(\a_3,\theta) = g_{\tau\tau} g_{xx}$, where eventually only the values 
$\th=0$ and $\th=\pi/2$ are allowed. 
Following standard procedures one can find integral expressions for
the separation of the two sources and the potential as function of
the maximal distance of the string from the boundary.
For small separations
we find, as expected, a Coulomb potential $V_{q\bar{q}} \sim -1/L$.
However, for larger values of $L$ we have to distinguish between the cases
of negative and zero $c$. 
Let us begin with the $c<0$ case: as we increase $L$ the energy
increases until it vanishes at a certain maximal length.
Beyond that there doesn't exist a geodesic connecting the two sources,
the configuration will be that of two disconnected straight strings
and the potential is zero.
We can interpret this as complete screening with the maximum length 
given by 
\be
L_{\rm max} = \left\{\begin{array}{ll}
\frac{\pi}{M_{\rm gap}}  \ , & \th=0\ , \\
\frac{\pi}{2 M_{\rm gap}} \ , & \th={\pi\ov 2} \ ,
\end{array} 
\right. 
\ee
where the mass gap $M_{\rm gap}$ is given by \eqn{mmggp}.
This behaviour of Wilson loops is precisely the one found in the
context of the Coulomb branch of theories with sixteen supercharges
\cite{brand1} for D3-branes distributed uniformly over a disc.

For $c=0$ the behaviour is quite different. There doesn't exist
a finite screening length and the potential vanishes only at an
infinite separation.
For a short separation the potential exhibits the usual Coulombic behaviour,
but we can also work out the behaviour for a very large separation.
The result depends on the direction on the sphere:
for $\theta=0$ we find again $V_{q\bar{q}} \sim -1/L$
which is reminiscent of a conformal theory, although the metric is not
that for the $AdS_5$ space-time. 
For $\theta = \pi/2$ the potential is screened, but it still exhibits a 
power-law behaviour as 
$V_{q\bar{q}} \sim -1/L^{2}$. This is consistent with the fact 
that there is a ring-type singularity for the metric at $\th=\pi/2$ as for 
the enhancon in \cite{jpp} and for a uniform distribution of D3-branes in 
a ring in 
\cite{sfe1}. For the latter case the Wilson loop was analyzed in \cite{brand1}
and also exhibits a $1/L^2$ behaviour for large separations.

Note also that, for the qualitative features that we have presented for the 
Wilson loops the value of the 10-dim dilaton computed in \cite{PW2} plays no 
r\^ole except for the case with $c=0$ and $\th=\pi/2$. Then the inclusion
of the dilaton factor is indeed 
necessary in order to obtain the $1/L^2$ behaviour that we have mentioned.

\section{Concluding remarks}

The 5-dim backgrounds that we described have naked singularities
in the interior (IR in field theory). Hence, the question arises
whether the physics or more specifically string theory is singular or not
in such backgrounds.
Well known examples of singular geometries that are resolved by
string theory are orbifolds, orientifolds and conifolds.
More recently singular geometries appeared in the supergravity duals of
non-conformal theories. Examples are dilatonic branes, backgrounds
dual to theories with sixteen supercharges on the Coulomb branch,
for which the singularity is the source of a distribution of
branes,
and duals of theories with eight supercharges where the singularity
is removed by a mechanism explained in \cite{jpp}.
We think that our geometries are similar in nature to the examples
of the Coulomb branch with sixteen supercharges.

Furthermore, the distinction between good and bad singularities
should be consistent with the AdS/CFT correspondence. This means
that only a well-defined deformation and vacuum-state in field theory
is dual to a geometry with a good singularity. We have encountered this issue
in our paper. For $c >0$ we seemingly try to give a vev to a massive
scalar field in the field theory. However, since we have turned on a non-zero
source (mass term) for these scalars, its vev has to be fixed uniquely.
Therefore, the case $c>0$ should correspond to a bad singularity,
and $c \leq 0$, on the other hand, should be an acceptable one.

Unfortunately, we do not know enough about string theory in such backgrounds
to answer the question whether the singularity is physical or not.
Instead, we can ask a somewhat simpler question: is the propagation
of a quantum test particle well defined in the presence of the singularity?
This criterion \cite{wald,homa,ishibashi} 
is identical to finding a unique
self-adjoint  extension of the wave-operator.
This can be quickly answered by looking at the two solutions of the
wave-operator locally near the singularity. The relevant part
of the wave equation is
\be
\frac{d}{dr} \left( \sqrt{-g} g^{rr} \frac{d\psi}{dr}\right) =0
\ee
and the norm is $\frac{q^2}{2} \int \sqrt{-g} g^{tt} \psi^\dagger \psi +
\frac{1}{2}\int \sqrt{-g} g^{ij} D_i\psi^\dagger D_j \psi$.
This is the Sobolev norm and it is bounded from above by a constant
times the energy of the fluctuation \cite{ishibashi}.
This is a physical sensible
norm because it guarantees that the backreaction of the fluctuation
is small.

For $c \leq 0$ only one solution is normalizable and the non-normalizable
is discarded. 
Therefore, there exists a unique self-adjoint extension
and the singularity is wave-regular \cite{ishibashi}, in accord with our
expectations. This is in accordance with the fact that, in these cases the 
singularity is null (as it occurs at $z=\infty$) and the evolution of 
initial date is the ordinary Cauchy which is unique.
However, for $c>0$, the metric near $z=z_0$ is approximated by 
\be
ds^2 = \tilde{x}^{1/3} (dx_{||}^2 + d\tilde{x}^2)\ ,
\ee
where $\tilde{x}=z_0-z \to 0^+$. 
Now both solutions $\psi \sim 1~,~\tilde{x}^{1/2}$
are normalizable and the singularity is wave-singular. This is consistent
with field theory expectations and a different criterion presented in
\cite{gubser}.
Actually, there is a subtlety in the choice of the norm of the wavefunction.
As refs. \cite{ishibashi,wald} 
we use the Sobolev norm which contains derivatives of
the wavefunction, whereas in \cite{homa} a norm without derivatives is
used. This distinction is important for 
example in the case of the negative mass Schwarzschild black hole:
using the Sobolev norm it is wave-regular, but it 
is singular in the convention of \cite{homa}.
We have also checked the criterion using the Sobolev norm in all backgrounds
corresponding to continuous D3-, M2- and M5-brane distributions describing 
states in Coulomb branch of the corresponding supersymmetric field theories.
Referring to the conventions in \cite{bakas1}, 
the naked curvature singularities occuring close to the brane distributions
are physical (unphysical) for $n\ge 2$ ($n=1$). This is in accordance with 
field theory expectations since in the unphysical cases the brane
distributions contain a component corresponding to negative tension branes
\cite{fgpw2,cglp}.
The criterion using the Sobolev norm is in accord
with the criterion of \cite{gubser} for all cases we are aware of,
but it seems to us that it is more natural to impose. 
There seems to be a consistent picture emerging using the simple
criterion of wave-regularity to distinguish acceptable singularities
from bad ones.
Evidently a better understanding of string theory in such backgrounds
is desireable, and in hindsight it is interesting to note
that this is one of the few examples in the AdS/CFT correspondence where
the field theory side can teach us something about the associated 
string theory.

\section*{Acknowledgements}

A.B. and K.S. would like to thank the Erwin Schr\"odinger Institute in Vienna
and the CIT-USC center for Theoretical Physics at USC in Los Angeles, 
respectively,
for hospitality and generous financial support during the final stages of 
this research, as well as for the opportunity to present these results in 
seminars prior to submitting the paper at hep-th.
Also A.B. thanks S.J. Rey for extensive
discussions.

\end{document}